\documentclass[conference,a4paper]{IEEEtran}
\usepackage{threeparttable}
\usepackage{amsmath,amssymb,epsfig,authblk,cite,booktabs}
\usepackage{comment}
\usepackage{color,hyperref}
\hypersetup{
    colorlinks,%
    citecolor=blue,%
    filecolor=blue,%
    linkcolor=blue,%
    urlcolor=blue
}
\definecolor{bluecolor}{rgb}{0,0.,1.}

\definecolor{redcolor}{rgb}{.7,0.,0.}

\newcommand{\vect}[1]{\boldsymbol{#1}}
\newcommand{\rev}[1]{{\color{red} #1}}
\newcommand{\revAK}[1]{{\color{blue} #1}}
\newcommand\norm[1]{\left\lVert#1\right\rVert}

\newcommand{\es}[1]{\begin{equation}\begin{split}#1\end{split}\end{equation}}

\newcommand{\argmax}{\operatornamewithlimits{argmax}}

\newcommand{\squeezeup}{\vspace{-1mm}}
\newcommand{\squeezeupone}{\vspace{-1mm}}
\newcommand{\squeezeuphalf}{\vspace{-0.3mm}}

\newcommand{\PP}{\mathbb{P}}

\begin{document}

\title{WiLAD: Wireless Localisation through \\Anomaly Detection}

\author[1,2]{Cam Ly Nguyen}
\author[3]{Aftab Khan}
\affil[1]{Network System Laboratory, Corporate Research \& Development Center, Toshiba Corporation}
\affil[2]{The University of Tokyo}
\affil[3]{Toshiba Telecommunications Research Laboratory}
\maketitle

\begin{abstract}

We propose a new approach towards RSS (Received Signal Strength) based wireless localisation for scenarios where, instead of absolute positioning of an object, only the information whether an object is \emph{inside} or \emph{outside} of a specific area is required.
This is motivated through a number of applications including, but not limited to, \emph{a) security:} detecting whether an object is removed from a secure location, \emph{b) wireless sensor networks:} detecting sensor movements outside of a network area, and \emph{c) computational behaviour analytics:} detecting customers leaving a retail store.
The result of such detection systems can naturally be utilised in building a higher level contextual understanding of a system or user behaviours. 
We use a supervised learning method to overcome issues related to RSS based localisation systems including multipath fading, shadowing, and incorrect model parameters (as in unsupervised methods). 
Moreover, to reduce the cost of collecting training data, we employ a detection method called One-Class SVM (OC-SVM) which requires only one class of data (positive data, or target class data) for training. 
We derive a mathematical approximation of accuracy which utilises the characteristics of wireless signals as well as OC-SVM. 
Based on this we then propose a novel mathematical formula to find optimal placement of devices. 
This enables us to optimize the placement without performing any costly experiments or simulations. 
We validate our proposed mathematical framework based on simulated and real experiments.

\emph{Keywords}
Wireless localisation, one-class classification, placement optimization, anomaly detection

\end{abstract}

\squeezeup\squeezeup\section{Introduction} 
\label{sec:intro}
\squeezeup
Wireless localisation, which has been of great interest over the past few years \cite{liu2007survey}, refers to extracting geo-location information of an object based on its wireless signals to multiple known devices. 
There are numerous important applications, particularly industrial applications, commercial environments, public safety settings, everyday life and defence/security systems \cite{stella2012rf}.
Solutions for deriving the location information can be categorized into two groups as unsupervised methods and supervised methods \cite{liu2007survey}.   
Unsupervised methods also known as triangulation methods estimate the distance from a number of known devices (anchors) and multilaterate the location of target objects \cite{savvides2001dynamic}. 
These methods are subject to errors that are caused by various factors including noise, multipath fading, shadowing and non line-of-sight (NLoS).
Moreover these are sometimes costly and time-consuming since model parameters need to be adjusted for specific environments. 
On the other hand supervised methods such as fingerprinting compare signal features to a pre-generated database in order to identify the most likely location of target objects \cite{rai2012zee}. 
Receive Signal Strength (RSS)-based location fingerprinting is commonly used for this method. 
There are various fingerprinting-based localisation algorithms such as probabilistic methods \cite{kontkanen2004topics}, $k$-nearest-neighbor (kNN), neural networks, support vector machine (SVM)\cite{wu2004wlan}, and smallest M-vertex polygon (SMP)\cite{prasithsangaree2002indoor}.
These methods often perform better than unsupervised methods, however they are computationally expensive and time consuming since signal fingerprints are required to be collected in advance.

These localisation methods provide absolute positions of target objects. 
However in some applications, the absolute positions are not always required. 
For example, in some scenarios a target object is only required to be detected whether it is \emph{inside} or \emph{outside} of a specific place. 
In security, it is crucial to detect whether an object (for example an object that can transmit wireless signals such as a smartphone, a tablet) is removed from a secure location. 
In the context of customer analytics (for example in retail), the main interests are in the number of people entering or leaving the store and the time they spend purchasing/viewing products. 
Such analytics can be enabled assuming that people have access to wireless devices such as smart phones.
With pre-designated zones in the store, a more thorough and complete understanding of consumer behaviour can be established. 
For example, analysis of customers entering and spending time at a particular section can be made where a new product has been recently launched.
Such localisation systems can also find an application in the sensor networks domain where a sensor node is required to be detected if it goes \emph{out} of its specific (i.e. usual) area. 
The information can be used such as to detect when some phenomenon happened (earth quake, landslide)....
Such kind of applications motivate us to define and develop a new class of localisation, WiLAD (Wireless Localisation through Anomaly Detection) which classifies the target object into two types of area: \emph{target area} (i.e. inside) and \emph{non-target area} (i.e. outside).
Due to the fact that the non-target area can be too large making data collection practically infeasible, collecting training data in only the target area, which is normally small, can significantly reduce data collection costs compared with conventional fingerprinting methods in which training data for all classes is required. Therefore, these methods cannot be directly utilised in our localisation system which requires only training data in one class. Besides, other information such as locations of anchors and model parameters (particularly needed in the case of unsupervised methods), are also not required, thus deployment requirements are minimal.
 
In order to identify objects of a specific class amongst all objects, we approach this with a one-class (or unary) classification mechanism. This is performed by learning from a training set containing only the objects of that class \cite{khan2014one}. 
Among one-class classification methods, one-class support vector machine (OC-SVM) is known to outperform other methods in several datasets \cite{khan2014one}. 
Therefore we employ OC-SVM in our localisation system. 
To the best of our knowledge, this is the first such attempt to perform wireless localisation. 


Besides, other works related to improving localisation accuracy, such as localisation accuracy estimation \cite{stella2012rf}, anchor placement optimization \cite{farkas2015placement, baala2009impact} has also attracted significant attention in recent times. 
For example, \cite{stella2012rf} derive the bound of localisation accuracy for RSS measurements. 
This gives a useful insight in to localisation performance and deployment issues of a localisation system, which could help in designing an efficient localisation system. 
In \cite{farkas2015placement, baala2009impact}, the authors set out to find the optimal number and placement of the anchor nodes in a given area for improving localisation accuracy. 
These methods rely on real experiments or simulations for specific environment (specifically requiring experiments or simulations for each of the given areas) thus raising both the cost and complexity.    
In this paper, we derive approximation formulation of accuracy which utilizes the characteristics of wireless signals as well as OC-SVM. 
Moreover, based on the formulation we then propose a novel mathematical framework to find the optimal placement of devices (anchor devices as well as target areas for target objects). 
This mathematical formulation enables us to optimize device placement without performing any costly experiment or simulation.

Our main contributions are summarised as following: 
\begin{itemize}
\item Propose a method to estimate the detection accuracy.
\item Propose methods to improve the accuracy including a novel method for optimizing placement of devices.
\item Validate proposed methods via numerical simulations as well as real experiments.
\end{itemize}


\squeezeup\squeezeup\section{Preliminaries}
\label{sec:pre}
\squeezeup\subsection{System model}
\squeezeup
Consider a wireless network system of $k$ anchors (here referred to as access points or APs) positioning at $\vect{a}_1, \vect{a}_2, ..., \vect{a}_k$, and a target object normally moving or staying around an area called \emph{target area}. 
The target object is equipped with a radio transceiver, and broadcasts beacon signals at a set interval of time. 
Each AP then receives the signals and retrieves RSSIs (Received Signal Strength Indicators), followed by sending values of RSSIs to a backhaul server. 
The server then uses collected RSSIs to determine whether the target object stays inside its target area or not (i.e. non-target area).

\squeezeup\squeezeuphalf\subsection{Propagation Models}
\label{sec:propagation}
\squeezeup

RSSI $r_{\vect{t}, i}$ between a target object positioning at $\vect{t}$ and the $i$-th AP positioning at $\vect{a}_i$ is related through the Friis equation (in dBm) \cite{bose2007practical}. 
\squeezeupone\es{
r_{\vect{t}, i}= P_T - 10 \eta \lg {\norm{\vect{t}- \vect{a}_i}}  + \mathcal{X}
\label{rssi-distance1}
}
where $\eta$ and $P_T$which are constants, are the path loss exponent and the transmit power respectively, $\mathcal{X}$ is a random variable characterising the effects due to multipath fading and noisy measurements. $\norm{\vect{x}}$ is Euclidean norm of a vector $\vect{x}$, thus 
$ \norm{\vect{x}- \vect{y}}$ is the distance between two positions $ \vect{x}$ and $\vect{y}$.
In this paper we denote $\log_{10} x$ as $\lg x$ for simplicity.
The signal fluctuations $\mathcal{X}$ due to multipath fading and noise depend on  the wireless propagation environment. 
For example, the long-term signal variation is known to follow the Log-normal distribution, whereas the short-term signal variation can be described by several other distributions such as Hoyt, Rayleigh, Rice, Nakagami-m, and Weibull. 


\squeezeup\squeezeuphalf\subsection{One-class support vector machine (OC-SVM)}
\label{sec:OC}
\squeezeup
OC-SVM (a particular type of supervised learning) tries to identify objects of a specific class amongst all objects, by learning from a training set containing only the objects of that class.
We briefly introduce OC-SVM \cite{scholkopf2001estimating} as follows. 
Suppose the training target class is ${\vect{\hat{r}}_1, \vect{\hat{r}}_2,..., \vect{\hat{r}}_s }$, where $ \vect{\hat{r}}_j \in \mathcal{R}^k, \forall j \in [1,s]$. 
In the input space, OC-SVM aims to determine a hyperplane to separate the target class and the origin of the input space with the maximum margin:
\squeezeup
\begin{equation}
\begin{aligned}
& \underset{}{\text{min}}
&&  \frac{1}{2} \norm{\vect{w}}^2 -\rho + \frac{1}{\vartheta \cdot s} \sum_{1 \leq j \leq s} \xi_j  \\
& \text{s.t.} &&  \vect{w} \cdot \vect{\hat{r}}_j \geq \rho - \xi_j \\
& &&   \xi_j \geq 0, \quad \forall j
\end{aligned}
\label{oc-svm}
\end{equation}
where parameter $\vartheta \in(0,1)$ is used to trade off the sphere volume and the errors $\sum_{1 \leq j \leq s} \xi_j $, $s$ is the size of the training data.
For a test sample $\vect{\hat{r}}_{\vect{t}}$ if
\squeezeupone\es{
 \vect{w} \cdot \vect{\hat{r}}_{\vect{t}} \geq \rho,  
\label{classify}
}
it is classified into the target class, otherwise, it belongs to the non-target class.
In practice,  $\vartheta$ is automatically calculated if provided the fraction of training error (called $\nu$).  
The inner product is normally calculated using a kernel. The Radial basis function kernel, also called the RBF kernel, or Gaussian kernel is widely used, which is defined as follows. 
\squeezeupone\es{
\vect{x}\cdot \vect{y}= \exp{(-\gamma \norm{\vect{x}- \vect{y}}^2) }
} 
where $\gamma$ is a constant. The kernel is the indicator of similarity between two vectors $\vect{x}$ and $\vect{y}$. 

\squeezeup\squeezeup\section{localisation anomaly detection method}
\label{sec:lads}
\squeezeup

Our proposed framework for anomaly detection in localisation system has two main phases: training phase and decision phase. 
In the training phase, data in the target class collected beforehand is used to train an OC-SVM.
In the detection phase, the trained model is used to determine whether the target object is inside the target area or outside using the data collected in real-time. 

To improve the accuracy, before passing to the OC-SVM, we perform feature extraction as follows.

\squeezeup\squeezeuphalf\subsection{Feature Extraction}
\squeezeup
As described in Section \ref{sec:propagation}, a single signal fluctuation normally follows a non-Gaussian distribution, in which, in extreme cases it is possible that the absolute value of random variable $\mathcal{X}$ becomes very large, i.e., RSSI between nodes is small even when their distance is close. 
Such fluctuations can have a significant effect on the detection accuracy. 
Therefore to improve detection accuracy, we average $N$ successive RSSI values between the target object an each APs, in which $N$ can be empirically selected depending on the applications. 
Therefore, the availability of multiple independent RSSI measurements enables the use of the Center Limit Theorem (CLT), and thus the modelling of fluctuation by a Gaussian distribution. 
The averaged RSSIs between the target object positioning at $\vect{t}$ and the $i$-th APs positioning at $\vect{a}_i$ follows:
\squeezeupone\es{
\bar{r}_{\vect{t}, i}= P_T - 10 \eta \lg {\norm{\vect{t}- \vect{a}_i}}  + X,
\label{rssi-distance}
}
where $X$ is a random variable (with a Gaussian distribution).


Secondly, in order to achieve a generalized applicability and a scalable method, we standardize our averaged data to minimize cross-environmental RSSI magnitude variance. 
Namely, in the training phase each averaged RSSI is subtracted by the mean from each feature type, then divided by its standard deviation. 
On the other hand, in the detection phase, the averaged RSSIs is subtracted by the mean from the corresponding features in training data.

\squeezeup\squeezeuphalf\subsection{Parameter settings for OC-SVM}
\squeezeup

To enhance the system accuracy, it is fundamental to choose appropriate parameters for the OC-SVM. 
While in binary SVM, the training data in both classes are available thus the parameters can be optimized using such as cross validation, in OC-SVM the parameters are difficult to be optimized since data in non-target class is unavailable.
The RBF's parameter $\gamma$ is therefore set to be at its default value, i.e., $\gamma= 1/k$, where $k$ is the number of features which is equal to number of APs. 
This is because, Section \ref{sec:prob} shows that the value of  $\gamma$ does not effect the OC-SVM strongly if the data is standardized.

\squeezeup\squeezeup\section{Probability of successful detection}
\label{sec:prob}
\squeezeup
In this section we propose a mathematical formulation for estimating the accuracy under some assumptions, aimed at providing meaningful insights towards achieving optimal accuracy. 
Given the fraction of training error $\nu$, the accuracy is related to the probability of successful detection (here called \emph{detection rate}) when the target node goes outside of its target area. 

\squeezeup\squeezeuphalf\subsection{Formulation}
\label{sec:form}
\squeezeup
For simplicity, we propose a method for calculating the detection rate under the following assumptions:
Firstly, we assume that the target area is small, so we can consider that it is a point positioning at $\vect{t}_{in}$. 
Note that if the target is not that small, we can approximately consider $\vect{t}_{in}$ as the middle point of the target area. 
For instance, $\vect{t}_{in}$ is illustrated by symbol X in Fig. \ref{fig:layout} which is the middle point of a particular target area.
Secondly, due to the averaging process described in section \ref{sec:lads}, fluctuation of each averaged RSSI can be assumed as following Gaussian distribution with $0$ mean, $\sigma^2$ variance, namely, $X \sim \mathcal{N} (0, \sigma^2)$.
Due to Equation \eqref{rssi-distance}, the averaged RSSI between target object positioning at $\vect{t}_{in}$ and the $i$-th AP follows $ \mathcal{N} (P_T- 10\eta\lg \norm{\vect{t}_{in}- \vect{a}_i} , \sigma^2)$. 
Thus the value of the $i$-th feature corresponding to the $j$-th training data is 
\squeezeupone\es{
\hat{r}_{j, i} = \frac{\bar{r}_{j,i}- P_T+ 10\eta\lg  \norm{\vect{t}_{in}- \vect{a}_i}}{\sigma}= \frac{X}{\sigma}.
\label{normalize}
} 
Hence each training vector $\hat{\vect{r}}_j$ consists of $k$ components following $\mathcal{N} (0, 1)$, where $k$ is the number of APs. 
We then estimate the margin of an OC-SVM trained by training data $\vect{\hat{r}}_1,...,\vect{\hat{r}}_s $. 
The first constraint of OC-SVM given by Formula \eqref{oc-svm} can be written as follows.
\squeezeupone\es{
 \exp{(-\gamma \norm{\vect{w}- \hat{\vect{r}}_j }^2)} \geq \rho- \xi_j \\
\Leftrightarrow \norm{\vect{w}- \hat{\vect{r}}_j }^2  \leq -\frac{\ln(\rho-\xi_j)}{\gamma}
\label{constraint}
}
Since the objective of an OC-SVM is to minimize the sum $\frac{1}{2} \norm{\vect{w}}^2 -\rho + \frac{1}{\vartheta \cdot s} \sum_{1 \leq j \leq s} \xi_j $, namely approximately minimize $ \norm{\vect{w}}$ and $\xi_j- \rho$. Thus $\norm{\vect{w}- \hat{\vect{r}}_j }^2$ (the left side of \eqref{constraint})  and $\norm{\vect{w}}$ should take small values. 
Moreover the average of training data $\hat{\vect{r}}$ is $\vect{0}$, consequently $\vect{w}$ is approximately also $\vect{0}$. 

Substituting $\vect{w}=  \vect{0} $ in Equation \eqref{classify}, a vector 
$\vect{\hat{r}}_{\vect{t}}$ can then be classified as in the target class if:
\squeezeupone\es{
 \exp{(-\gamma \norm{\vect{w}- \hat{\vect{r}}_{\vect{t}}}^2)} \geq \rho \\
\Leftrightarrow  \norm{\hat{\vect{r}}_{\vect{t}}}^2    \leq \delta,
\label{decision}
} 
where $\delta= -\ln(\rho)/\gamma$. 

Since vector $\hat{\vect{r}}_j$ has $k$ components, in which each component $\hat{r}_{j,i} $ follows $\mathcal{N} (0, 1)$ and are independent to each other, thus $\norm{\hat{\vect{r}}_j }^2$ follows chi-squared distribution with $k$ degrees of freedom. 
As fraction of training error is $\nu$, there is $1-\nu$ fraction of training data satisfying Equation \eqref{decision}. Thus, 
\squeezeupone\es{
\delta= F_{\chi2}^{-1} (1-\nu)
\label{delta}
}
where $F_{\chi2}(x)$ is the cumulative distribution function (CDF) of variable $\chi2$ following chi-squared distribution with $k$ degrees of freedom, evaluated at $x$, and $F_{\chi2}^{-1}(x)$ is its inverse function. 
Equation \eqref{delta} shows that the value of $\delta$ only depends on $\nu$, thus is a constant.

We now calculate the \emph{detection rate} of a specific position $\vect{t}$, which is the probability that the trained OC-SVM classifies a vector  $\vect{\hat{r}}_{\vect{t}} $ as non-target class when the target object positioned at $\vect{t}$ is outside its target area.
Averaged RSSI between $\vect{t}$ and AP $\vect{a}_i$ can be described as follows.
\squeezeupone\es{
\bar{r}_{\vect{t}, i} = P_T- 10\eta\lg \norm{\vect{t}- \vect{a}_i} + X . 
}
Utilising Equation \eqref{normalize}, the standardized vector $\vect{\hat{r}_{\vect{t}}}$ has $i$-th component having the following value: 
\squeezeupone\es{
\hat{r}_{\vect{t}, i} &= \frac{\bar{r}_{\vect{t}, i} -P_T+ 10\eta\lg \norm{\vect{t}_{in}- \vect{a}_i}}{\sigma} \\
&=  \frac{10\eta}{\sigma} \lg {\frac{\norm{\vect{t}_{in}- \vect{a}_i}}{\norm{\vect{t}- \vect{a}_i}} }+ \frac{X}{\sigma},
}
which follows  $\mathcal{N}(  \frac{10\eta}{\sigma} \lg {\frac{\norm{\vect{t}_{in}- \vect{a}_i}}{\norm{\vect{t}- \vect{a}_i}} } , 1)$
Thus we have,
\squeezeupone\es{
\norm{\hat{\vect{r}}_{\vect{t}} }^2 = \sum_{i\in [1,k]}  (\frac{10\eta\lg {\frac{\norm{\vect{t}_{in}- \vect{a}_i}}{\norm{\vect{t}- \vect{a}_i}} }+ X}{\sigma})^2
}
which follows non-central chi-squared distribution with $k$ degrees of freedom and non-centrality parameter $\lambda_{\vect{t}}/\sigma^2$, where, 
\es{
\lambda_{\vect{t}}= \sum_{i\in [1,k]}  (10\eta \lg {\frac{\norm{\vect{t}_{in}- \vect{a}_i}}{\norm{\vect{t}- \vect{a}_i}}} )^2. 
\label{lambda}
}
The target object is classified as non-target area (cf. \eqref{decision}) iff:
$\norm{\hat{\vect{r}}_{\vect{t} }}^2  > \delta.$ 
Therefore, the detection rate, i.e., probability that the target object $t$ (called $R(t)$) is classified as non-target area is:
\es{
R(\vect{t}) &= \PP[\norm{\hat{\vect{r}}_{\vect{t}} }^2 > \delta] \\
&= 1- P(\delta; k, \lambda_{\vect{t}}/\sigma^2)\\
&= Q_{k/2} (\sqrt{\lambda_{\vect{t}}}/\sigma, \sqrt{\delta}),
\label{detectRate}
}
where $\delta$ can be calculated using Equation \eqref{delta},
 $P(\delta; k, \lambda_{\vect{t}}/\sigma^2)$ is the CDF evaluated at $\delta$, of a random variable following non-central chi-squared distribution centering at $\lambda_{\vect{t}}/\sigma^2$ and having $k$- degrees of freedom. 
This CDF can be calculated by Marcum Q-function $Q_{k/2} (\sqrt{\lambda_{\vect{t}}}/\sigma, \sqrt{\delta})$ which is proved to be monotonic \cite{sun2010monotonicity}. 
Moreover as $\delta$ is a constant, $R(\vect{t})$ is a monotonic function of $\sqrt{\lambda_{\vect{t}}}/\sigma$.

The detection rate of a domain $D$, which is the probability that a trained OC-SVM classifies the target object as non-target area, when the target object positioned at an arbitrary point inside domain $D$, is:
\squeezeupone\es{
R(D) = \frac{1}{V_D} \int_{D} R(\vect{t}) d\vect{t},
\label{detectRateWhole}
}
where $V_D$ is the volume of domain $D$.
\squeezeup\squeezeuphalf\subsection{Stability of the proposed formulation}
\label{sec:sim}
\squeezeup
In practice, the signal attenuation due to path loss and its fluctuations due to multipath fading are often more complicated than suggested by Formula \eqref{rssi-distance}.
To investigate the appropriateness of the proposed Equation \eqref{detectRate} as well as to analyse factors that affect the detection accuracy, we conduct Monte Carlo simulations under two different propagation models: One following Formula \eqref{rssi-distance} and the other following a more advanced propagation model described below.

\subsubsection{Advanced propagation model}
\label{sec:advance}
We simulate a propagation environment experiencing Rayleigh fading, non-singular path loss. The RSSI values $r$ under this propagation model are generated via: 
\squeezeupone\es{
r= P_T - 10  \lg (\epsilon + d^\eta) + \mathcal{X}
\label{rssi-distance-alt}
}
where $\epsilon>0$, $d$ is the distance between two wireless devices, and $\mathcal{X}$ is a random variable with density:
\squeezeupone\es{
f_{\mathcal{X}}(x) &= \lambda 10^{x/10} \exp\Big(-\lambda 10^{x/10} \Big) \frac{\ln 10}{10}
\label{pdf}}
Recent indoor measurements at $2.4$GHz \cite{howmany16} have confirmed the above model.
where $\lambda$ is a constant, and here we set $\lambda\!= \!0.561$ because in this case the mean of $\mathcal{X}$ is zero \cite{nguyen2017wireless}.
A meaningful correspondence between $\mathcal{X}$ and our simplified Gaussian approximation $X$ can be established ($\sigma = 5.57 $).

\subsubsection{Parameter settings}
Assuming that there are three APs located at positions having coordinates of $[0,0], [0,10], [10,0]$ (in meters). 
The $\vect{t}_{in}$ corresponding to the target area is set at $[5,5]$. 
The fraction of training error $\nu$ is set as $0.1$.
We calculate and compare the detection rate $R(\vect{t})$ at $20$ positions of $\vect{t}$ that are approximately $3$-$30$m from $\vect{t}_{in}$.

For each position of the target object, we generated $1000$ sets of data, in which each set contained $3$ RSSIs from the target object to three APs. 
In each random realisation and for each pair of target object and AP, RSSI is generated randomly under two propagation models given by \eqref{rssi-distance} utilising random variable $X\sim \mathcal{N}(0, \sigma^2) $ and \eqref{rssi-distance-alt} and the random variable $X$ with its probability density described by \eqref{pdf} and $\epsilon= 0.1$, and common parameters $\sigma= 0.57$, $\eta= 2$ and $P_T= -30 dBm$.
The detection rate $R(\vect{t})$ for each position $\vect{t}$ is the percentage of data classified as non-target area.

\subsubsection{Results}
\begin{figure}
\centering
\begin{tabular}{r  l}
\includegraphics[
width=0.48\linewidth]{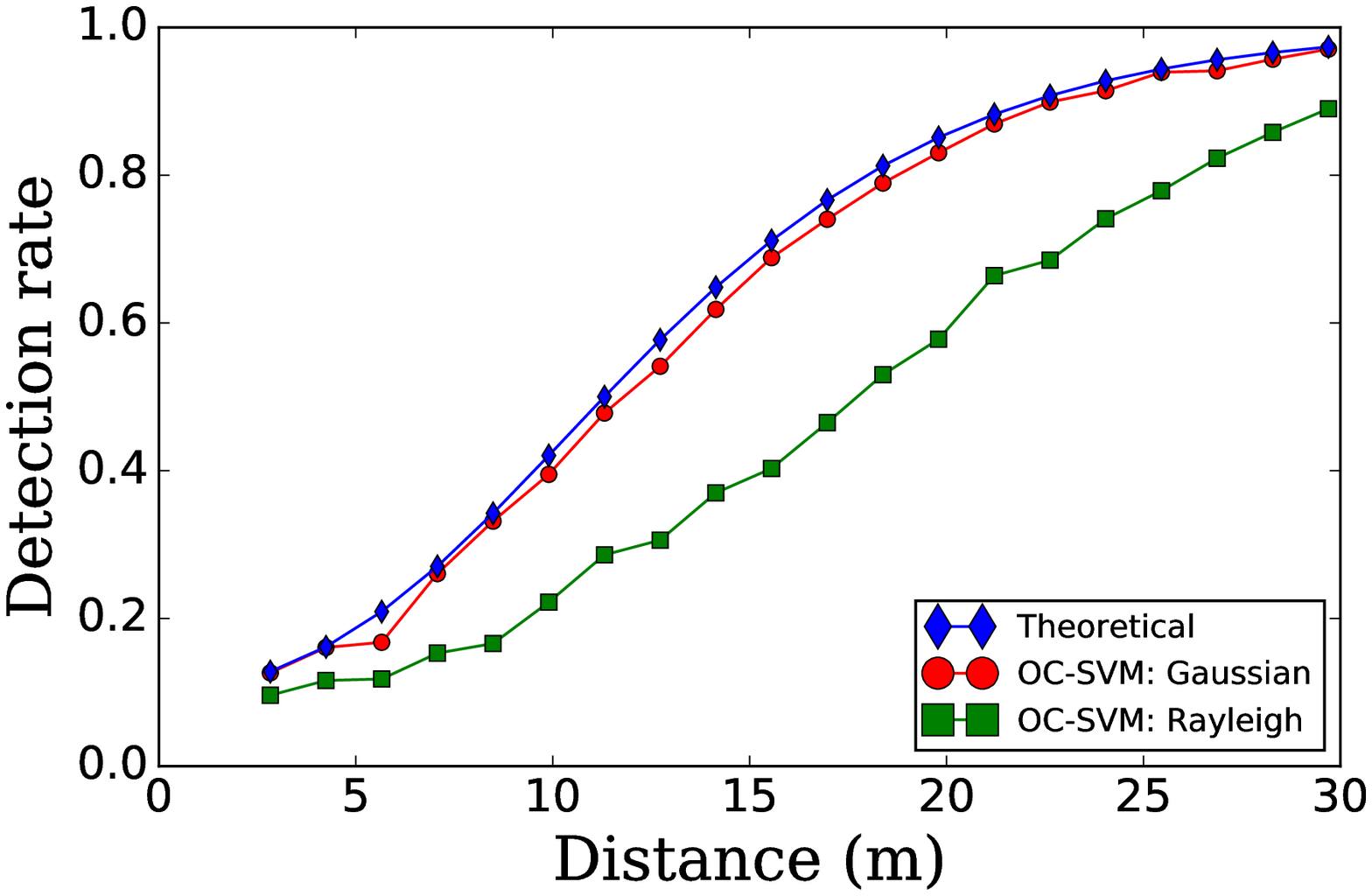}
&
\includegraphics[
width=0.48\linewidth]{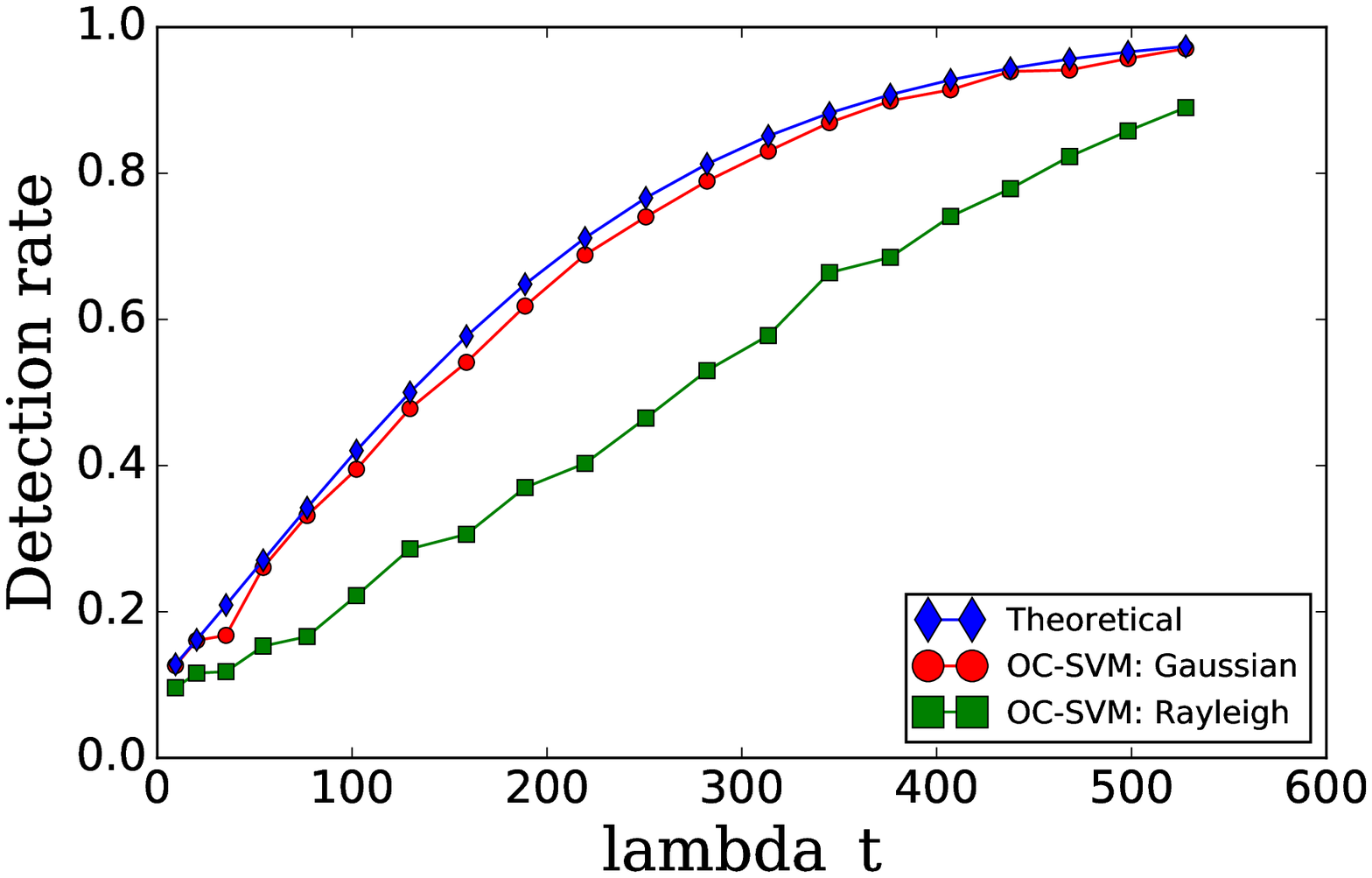}
\end{tabular}
\caption{Detection rate when the target object moved out of its target area: Blue diamonds describe the rate calculated by the proposed Equation \eqref{detectRate}, red circles and green square illustrate the detection rate achieved using simulation under the assumption that RSSI follows Equations \eqref{rssi-distance} and \eqref{rssi-distance-alt} respectively. X-axis corresponds to the distance from the target object to the target area in a) (the left figure), and to the $\lambda_t$ in b) (the right figure).}
\label{fig:sim_result}
\end{figure}

When the signal attenuation due to path loss follows Friis model described by Equation \eqref{rssi-distance}, and its fluctuations following Gaussian distribution, the detection rate $R(\vect{t})$ calculated by simulation are closed to the proposed formula \eqref{detectRate} (shown in Fig \ref{fig:sim_result}). 
It indicates that under the assumption that random variables $X$ follows Gaussian distribution, our proposed formula \eqref{detectRate} is accurate.
  
When RSSIs follow a different model (e.g. following \eqref{rssi-distance-alt}), detection rate $R(\vect{t})$ by simulations is not close to the proposed Equation \eqref{detectRate} as it has been derived under the assumption that RSSIs strictly follow Equation \eqref{rssi-distance} for simplicity.
On the other hand, Fig. \ref{fig:sim_result} b) verifies the our claim that the detection rate is a monotonic function of $\lambda_{\vect{t}}$ regardless of propagation models. 
It indicates that the accuracy (i.e. detection rate) can be optimized by maximizing $\lambda_{\vect{t}}$, which provides meaningful insights towards optimizing placement of APs, as well as target area of the object as discussed in Section \ref{sec:optPos}.

\squeezeup\squeezeup\section{Optimization methods}
\label{sec:opt}
\squeezeup
In this section, we propose two optimisation methods -- capable of improving the detection rate -- utilising our proposed solution in Equations \eqref{detectRate} and \eqref{detectRateWhole}. 
These equations illustrate that for any domain $D$ and any position of the APs, the detection rate $R(D)$ increases when $\sigma$ decreases. 
In this work, we hypothesise that $\sigma$ can be decreased by averaging successive RSSI, as also discussed in Section \ref{sec:reduce}. 
Moreover, we also propose a novel mathematical formulation to find optimal placement of  APs as well as target areas which maximizes the detection rate of a specific domain $D$, detailed in Section \ref{sec:optPos}. 
The proposed formulation is environment independent, i.e. it can be used in any environment, enabling us to establish optimized placement of APs/target-areas without performing any costly experiments.
\squeezeup\squeezeuphalf\subsection{Averaging}
\begin{figure}
\centering
\begin{tabular}{r  l}
\includegraphics[ 
width=0.48\linewidth]{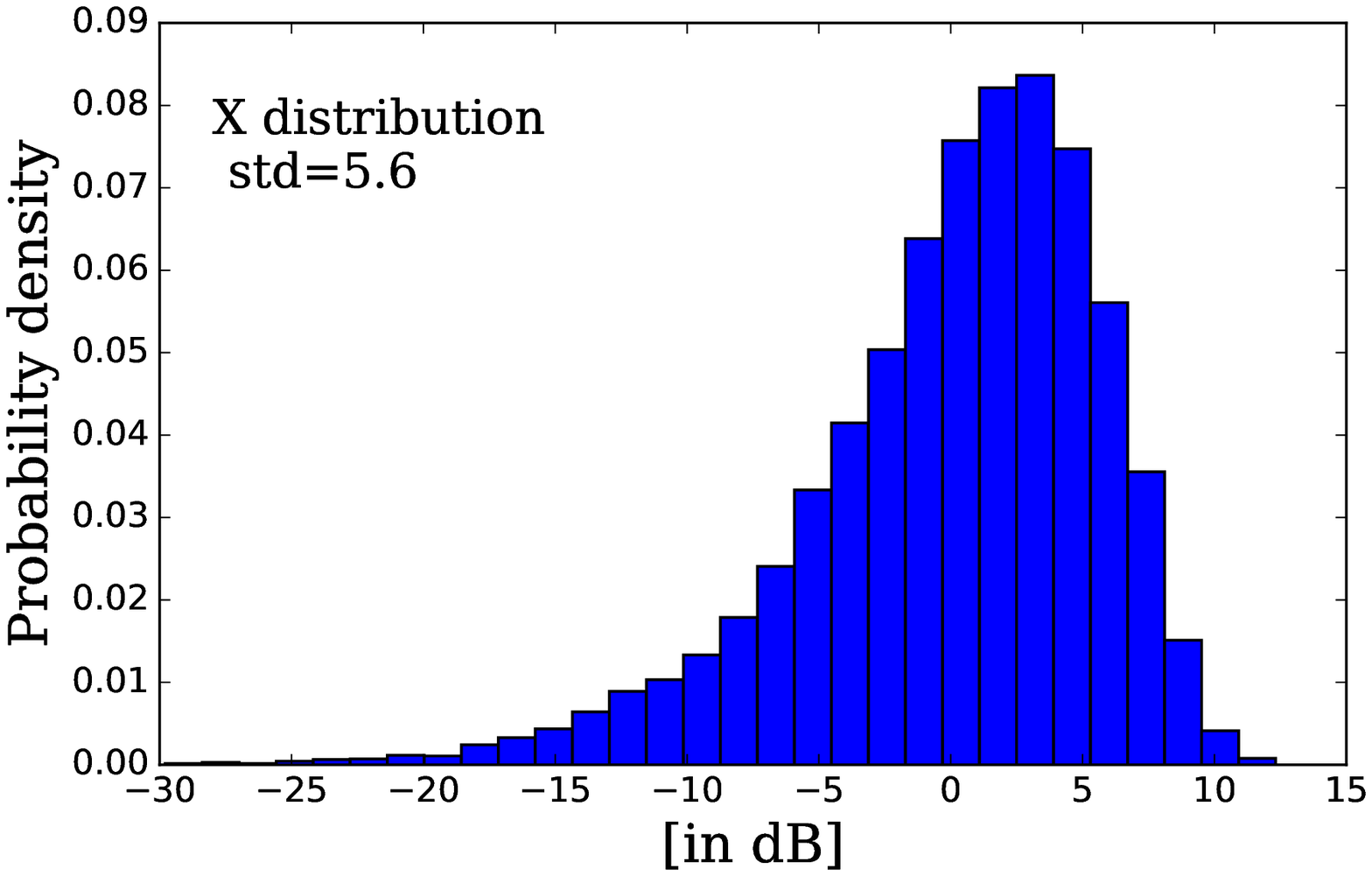}
&
\includegraphics[
width=0.48\linewidth]{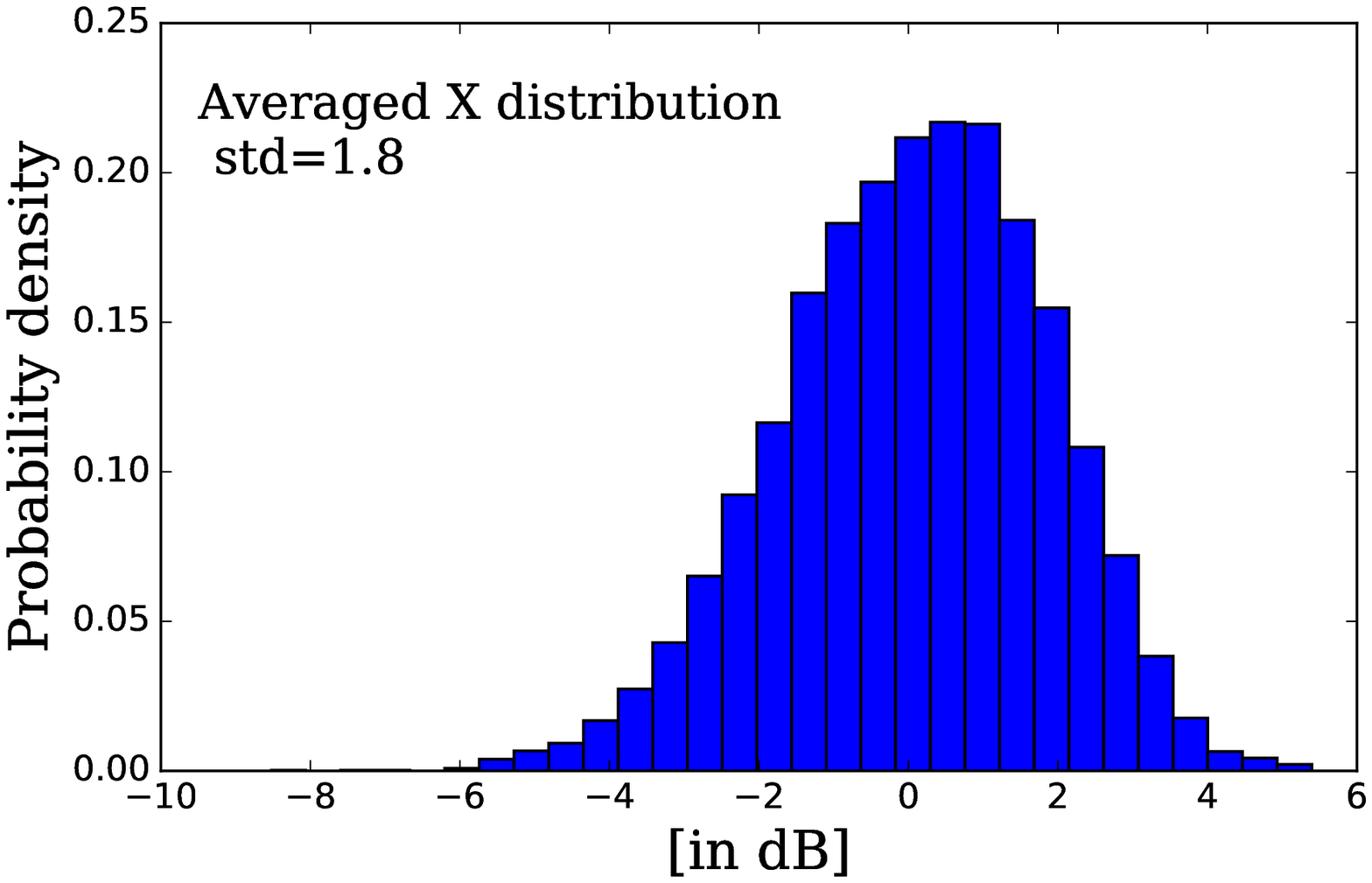}
\end{tabular}
\caption{ Probability density of the RSSI fluctuations due to multipath fading and noise ($\mathcal{X}$). a) Single RSSI (Left figure), b) averaged 5 successive RSSI (Right figure) }
\label{fig:noise}
\end{figure}

\squeezeup
\label{sec:reduce} 
Averaging successive RSSI can reduce the standard deviation of the signal resulting in less fluctuation and enabling an improved detection accuracy. 
To illustrate this, we pick signal fluctuations experiencing Rayleigh fading, namely probability density $\mathcal{X}$ following Equation \eqref{pdf}, for example. 
The probability density of $\mathcal{X}$ is illustrated by Fig. \ref{fig:noise} a), and its standard deviation is approximately $5.56$. 
Fig. \ref{fig:noise} b) shows the probability distribution of averaged five time series successive RSSIs and its standard deviation is now approximately $1.8$, which is much smaller than $5.56$, the standard deviation of the single signal fluctuation.

\squeezeup\squeezeuphalf\subsection{Device placement optimization}
\squeezeup
\label{sec:optPos}
In this section, we discuss the proposed solution to the optimization problem of placing APs/target-areas in order to maximize the detection rate.

\subsubsection{Problem definition}
In the context of our application case-study, a store would like to locate a set $A_k$ comprised of $k$ access points (APs) and a set $T_m$ consisting $m$ target objects. 
Each AP can choose its position from a set $A_K$ consisting $K$ candidate positions, and each target object can choose its target area from a set $T_M$ with $M$ candidate areas. 
Note that any two APs/objects cannot choose the same candidate position/area. 
The objective is to choose appropriate positions/areas for APs/objects to maximize the detection rate especially when a target object goes out of the store. 

\subsubsection{Proposed method}
We first define some symbols. 
Utilising a set  $A_k$ of APs, we denote $R(D | T_m, A_k) $ as the detection rate of domain $D$ given target areas $T_m$, which is the value we would like to maximize;  $R(\vect{t} | T_m, A_k) $ as the detection rate when any target object in $T_m$ reaches position $\vect{t}$; and $R(\vect{t} | \vect{t}_{in}, A_k) $ as the detection rate if the target object with its target area being $\vect{t}_{in}$ reaches the position $\vect{t}$.

We solve the problem under assumptions described in Section \ref{sec:prob} and that the store is separated by walls that absorb wireless signals. 
Therefore the detection rate $R(\vect{t} | T_m, A_k)$ at any position $\vect{t}\in D $ (where $D$ is the domain outside the store, see Fig. \ref{fig:layout}) is not smaller than the detection rate $R(\vect{t}_d)$ of the position $\vect{t}_d$ which is the position in the middle of the gate (see Fig. \ref{fig:layout}). 
We have:
\squeezeupone\es{
 R(\vect{t} | T_m, A_k) \geq  R(\vect{t}_d | T_m, A_k), \quad \forall \vect{t} \in D \\
\Rightarrow R(D | T_m, A_k) \geq  R(\vect{t}_d | T_m, A_k).
\label{maxD}
}
therefore maximizing the detection rate $R(D | T_m, A_k)$ is approximately maximizing the detection rate $ R(\vect{t}_d | T_m, A_k)$. 
Moreover, the store would like to detect if any target object goes outside, thus $ R(\vect{t}_d | T_m, A_k)$ can be defined as:
\squeezeupone\es{
 R(\vect{t}_d | T_m, A_k)= \min_{\vect{t}_{in} \in T_m}  R(\vect{t}_d | \vect{t}_{in}, A_k)
\label{maxT}
}
Consequently the objective of the problem is maximizing the right side of Equation \eqref{maxT}. 
Therefore, the objective of the problem can be written as follows:
\squeezeupone\es{
A_k^*, T_m^* = \argmax_{A_k \subset A_K, T_m \subset T_M} \min_{\vect{t}_{in} \in T_m} R(\vect{t}_d | \vect{t}_{in}, A_k)
\label{opt}
}

Since $R(\vect{t}_{d} | \vect{t}_{in}, A_k) $ is a monotonic function of $\lambda(\vect{t}_{d}, \vect{t}_{in}, A_k) /\sigma^2$ (see Section \ref{sec:prob}), where 
\squeezeupone\es{
\lambda(\vect{t}_{d}, \vect{t}_{in}, A_k)  = \sum_{\vect{a}\in A_k}  (10\eta \lg  {\frac{\norm{\vect{t}_{in}- \vect{a}}}{\norm{\vect{t}_d- \vect{a}}} })^2
\label{lambda}
}
$\eta$ is a constant reducing Equation \eqref{opt} to:
\squeezeupone\es{
A_k^*, T_m^* = \argmax_{A_k \subset A_K, T_m \subset T_M} \min_{\vect{t}_{in} \in T_m}  \sum_{a\in A_k}  ( \lg \frac{\norm{\vect{t}_{in}- \vect{a}}} {\norm{\vect{t}_d- \vect{a}} } )^2
\label{opt2}
}
Thus the optimal positions of APs and target areas can be calculated easily and efficiently -- without doing costly experiments or simulations -- in order to maximize the detection rate when an object is moved out of the store.

\squeezeup\squeezeup\section{Experimental investigation of WiLAD}
\label{sec:exp}
\squeezeup

In order to evaluate the performance of WiLAD in real environments and to validate our optimization methods described in Section \ref{sec:opt}, we performed experiments in a real store environment (see Fig. \ref{fig:exp}).

\squeezeup\squeezeuphalf\subsection{Experimental Setup}
\begin{figure}[t]
\begin{center}
\includegraphics[width=0.9\linewidth]{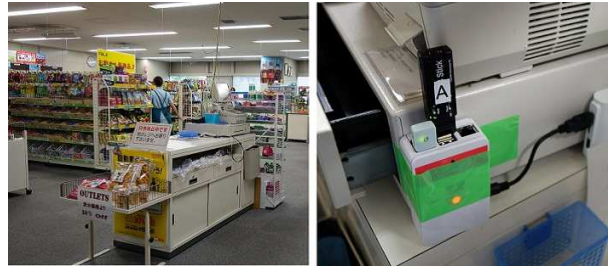}
\caption{Photos of the store, and a close-up of an RL7023 Stick-L as an AP.}
\label{fig:exp}
\end{center}
\end{figure}

\squeezeup
The experiments were conducted at a store in which the area inside and outside the store is approximately $120 m^2$ and $40 m^2$ respectively (area of Z5 in Fig. \ref{fig:layout}). 
The store is separated between inside and outside by concrete walls. 
Inside the store there are some obstacles such as goods shelves ($1.6m$ of height)
, tables ($0.8m$ of height; see Figs. \ref{fig:exp} and \ref{fig:layout}).
We used multiple Tessera RL$7023$ Stick-L acting as APs as well as target objects in the experiment, using $920MHz$ band. 
These IEEE$802.15.4$ standardized devices operate at $926.9MHz$ and house a patch antenna transmitting at $13dB$. 
Four APs were placed at positions described by red points labelled as A$1$ to A$3$ ($0.8m$ of height) and A$4$ ($2m$ of height) in Fig. \ref{fig:layout} .
Other three RL$7023$ Stick-L acting as target objects, could move around their target areas described by blue rectangles labelled as Z1 to Z3 in Fig. \ref{fig:layout}. 
During the experiment, the target objects broadcasted beacons every second; the APs after receiving the beacons and getting the RSSI would send it to a server for post-processing.  

To collect data, we first divided the space of the store into five zones, illustrated by Z1 to Z5 in Fig. \ref{fig:layout}. 
Z1 to Z3 (shown as the blue rectangles) are target areas corresponding to three target objects. 
Z4 is the remaining area inside the store and Z5 is the area outside of the store. 
We then installed wireless devices (RL$7023$ Stick-L) collecting 4 to 9 sets of data in each zone, and each set containing approximately $400$ subsets of data (where each subset contains four RSSIs from the target object to four APs). 
The experiment was conducted at different times of the day covering a range of business hours from less busy (few people in the store) to very busy (many people in the store).

\squeezeup\squeezeuphalf\subsection{Evaluation methodology}
\label{sec:eval}
\squeezeup
\subsubsection{Cross validation}
For each pair of target area and non-target area, we used a leave-one-out cross validation (LOOCV) scheme and calculated the evaluation value (i.e. detection rate or F-measure).
For instance, consider that Z1 is the target area, we used $7$ of the total $8$ sets of data collected at Z1 as the training data, and the remaining set as the test data (positive data), and also the data in non-target area as the test data (negative data).

\subsubsection{Detection rate and F-measure}  
Assuming the scenario described in Section \ref{sec:opt}, we estimate the detection rate showing the percentage of detections made by the OC-SVM when a target object goes out of the store, namely, $R(Z5)$. 
We calculate the detection rate as follows. 
For each target object (target area), similar to LOOCV described in the previous section, we use $S-1$ data sets ($S$ is the number of data set for the target area) as the training data. 
We then use the trained OC-SVM to calculate the percentage of successful detections when the target object stays in Z5 followed by repeating this calculation $S$ times for other target objects and averaging the results.
Detection rate is used in Experiment 1 below. On the other hand, F-measure \cite{powers2011evaluation} is used to evaluate the performance of WiLAD in Experiment 2 below.

\squeezeup\squeezeuphalf\subsection{Experiment 1}
\squeezeup
In order to validate our proposed optimization method for installation points given by Equation \eqref{opt2} in Section \ref{sec:opt}, we vary the value of $k\in [1,3], m \in[1,3]$ 
which are the number of APs and number of target objects that the store would like to setup, respectively. 
$k$ APs can choose their positions from $4$ candidates depicted by red points in Fig. \ref{fig:layout}. 
$m$ target objects can choose their areas from $3$ areas Z1, Z2, Z3 (namely $K= 4, M= 3$). 
Similar to the problem described in Section \ref{sec:opt}, the objective is to choose the best combination of $k$ AP positions and $m$ target areas that maximizes the detection rate when one of the target object goes outside the store.

To compare the solutions based on the proposed Equation \eqref{opt2} and our experimental solutions, for each values of $k,m$, we first list all feasible solutions, then sort the list using Equation \eqref{opt2} as well as based on the detection rate by experiment  described in Section \ref{sec:eval}. 
Here, we set a small fraction of training error ($\nu= 0.02$) to enlarge the sphere volume of the OC-SVM.
We then calculate the Pearson's correlation coefficient and its p-value between the two lists.

These results are shown in Table \ref{tab:result1}. 
For each pair $k,m$, we list the optimal solution based on the experimental setup followed by the number describing its order based on the theoretical representation (i.e., the proposed Equation \eqref{opt2}). 
Text in bold describe solutions that have the same order in both the experiments and the proposed formulation. 
For example, when $k=1$, and $m=2$, the best solution is A[1], Z[1,2] which means that the detection rate is maximum if the AP is set at A1, and two target objects are set at Z1 and Z2.  
It is ranked $1$ based on our proposed formula and the experimental evaluation. 
For all values of $k,m$, the correlation coefficient ranges from $0.60$ to $1.00$ with corresponding $p\leq 0.05$ in most cases, showing that our proposed approach is appropriate. 
Table \ref{tab:result1} also shows that
$67\%$ of the optimal solutions by the proposed formulation match the optimal solutions by experiments. 
In some cases, where the proposed formulation produces a different solution can be attributed to various environmental factors that RSSIs experience including multipath fading, shadowing, and NLoS. 
\begin{table}[t]
\renewcommand{\arraystretch}{1.3}
\caption{Experiments vs Theoretical results}
\label{tab:result1}
\centering
\scriptsize
\begin{tabular}{llcc}
\toprule
   $k, m$ &  \bfseries Solutions &  \bfseries Correlation  & \bfseries p-value\\
\midrule
 1, 1&  \textbf{A[3], Z[3]  \hspace{2mm} (1) }&  0.79& $2\mathrm{e}-3$ \\
 1, 2& \textbf{A[1], Z[1,2]  \hspace{2mm} (1)}& 0.66& 0.02\\
 1, 3&  A[4], Z[1,2,3] \hspace{2mm} (2) & 0.60 &0.40 \\
 2, 1 & \textbf{ A[3,4], Z[3] \hspace{2mm} (1)} & 0.92& $1\mathrm{e}-7$\\
 2, 2&  A[1, 3], Z[1, 3] \hspace{2mm} (2) &0.90&$3\mathrm{e}-7$ \\
 2, 3& \textbf{A[1, 3], Z[1, 2,3] \hspace{2mm} (1)}&0.89&0.02\\
 3, 1&   \textbf{A[1, 3, 4], Z[3] \hspace{2mm} (1)}  &0.76&$4\mathrm{e}-3$\\
 3, 2&  A[2, 3, 4], Z[2, 3] \hspace{2mm} (6)  &0.80&$2\mathrm{e}-3$\\
 3, 3&\textbf{A[1, 2, 3], Z[1, 2, 3] \hspace{2mm} (1)} &1.00&0.00 \\
\bottomrule
\end{tabular}
\end{table} 
\squeezeup\squeezeuphalf\subsection{Experiment 2}
\begin{figure}[t]
\begin{center}
\includegraphics[
width=0.9\linewidth]{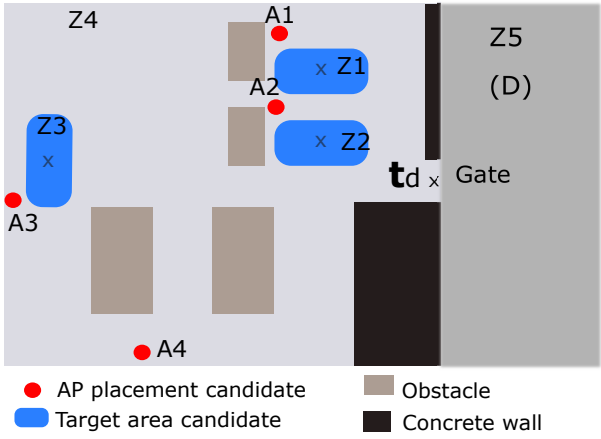}
\caption{Layout of the experiment.}
\label{fig:layout}
\end{center}
\end{figure}

\squeezeup
To evaluate the performance of the proposed WiLAD system in Section \ref{sec:lads} and to validate our approach described in Section \ref{sec:reduce}, we performed the experiment under the following scenario. 
There are three target objects with target areas namely Z1, Z2, Z3. 
Using three APs positioning at A1, A2, A3 (see Fig. \ref{fig:layout}), we are mainly interested in detecting whether a target object stays inside its target area or goes out of that area. 
Note that the positions of target areas as well as APs are chosen using the results of the previous experiments: optimal solution for $k=m=3$. 

For each target object, we define its non-target area as, 1) the remaining area of its target area located \emph{inside} the store (i.e. non-target area of the first object is Z2+Z3+Z4. 
The main purpose is to estimate the decision accuracy of WiLAD under the assumption that the object stays inside the store which is one of the \emph{non-target areas} and 2) the remaining area of its target area (i.e. which is the non-target area of the first object i.e., Z2+...+Z5). 
This is because, we are interested in estimating the decision accuracy of WiLAD under the assumption that the object stays inside the store or outside the store, called \emph{combined} non-target area. 
In each pair of target and non-target areas, we calculate the F-measure (see Section \ref{sec:eval}) using two type of data: 1) raw data (i.e. use single RSSI, namely set $N= 1$, where $N$ is number of data to be averaged, see Section \ref{sec:lads} ) and 2) averaged RSSI using $5$ successive data points (i.e. $N= 5$); here called \emph{raw data} and \emph{averaged data} respectively. 
We set the fraction of training error as $0.1$ (i.e. $\nu= 0.1$).

\begin{figure}[t]
\begin{center}
\includegraphics[ 
width=0.9\linewidth]{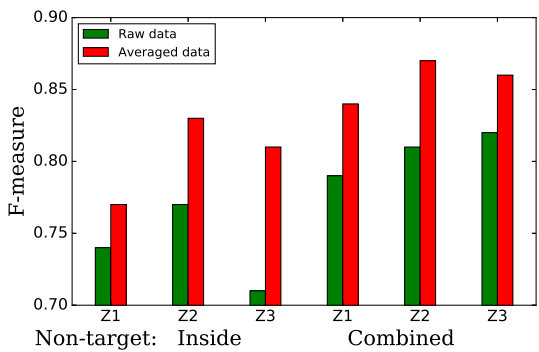}
\caption{F measure under different pair of target area- non-target area.}
\label{fig:result1}
\end{center}
\end{figure}

The mean F-measure depicted in Fig \ref{fig:result1} shows that the averaged data provides significantly better results compared with the raw data in every case (t-test, $p\leq 0.05$). 
The overall results achieved are always greater than $0.75$ illustrating a highly reliable system, further proving our arguments given in Section \ref{sec:reduce}. 

\squeezeup\squeezeup\section{Conclusions and Future Work}
\label{sec:con}
\squeezeup
In this paper, we proposed a new type of RSS (Received Signal Strength) based localisation method called WiLAD, and in particular addressing the problem of determining whether an object is inside its target area or not. 
Examples of such scenarios are commonly found in real life, for instance in security, in outlier detection of a wireless sensor network, or in customer analytics. 
We employed a one class classifier (OC-SVM) to classify an object in either target or non-target areas using its RSSIs to a number of known access points. 
We also derived an approximation formulation for estimating the accuracy and used it to derive a mathematical framework to optimize device placements. 
Finally, we validated our approach through experiments in a real store. 
Our results showed that $67\%$ of the optimal solutions by the proposed method match optimal solutions by experiments. 
Moreover, the achieved F-measures are always greater than $0.75$ illustraing a high reliability.


Despite such encouraging results, there is still much progress that can be made such as performing experimental evaluation in other indoor/outdoor environments, and utilising various wireless devices, transmission bands etc. 
Most of the proposed WiLAD framework has assumed that the wireless devices are deployed in line-of-sight (LoS) environments with perfect isotropic antennas. 
Generalising to non line-of-sight (NLoS) environments (e.g. multi-storey or multi-room building deployments) with anisotropic antennas can be a very interesting research extension.


\bibliographystyle{ieeetr}
\bibliography{mybib}

\begin{thebibliography}{10}

\bibitem{liu2007survey}
H.~Liu, H.~Darabi, P.~Banerjee, and J.~Liu, ``Survey of wireless indoor
  positioning techniques and systems,'' {\em IEEE Transactions on Systems, Man,
  and Cybernetics, Part C (Applications and Reviews)}, vol.~37, no.~6,
  pp.~1067--1080, 2007.

\bibitem{stella2012rf}
M.~Stella, M.~Russo, and D.~Begusic, ``Rf localization in indoor environment,''
  {\em Radioengineering}, vol.~21, no.~2, pp.~557--567, 2012.

\bibitem{savvides2001dynamic}
A.~Savvides, C.-C. Han, and M.~B. Strivastava, ``Dynamic fine-grained
  localization in ad-hoc networks of sensors,'' in {\em Proceedings of the 7th
  annual international conference on Mobile computing and networking},
  pp.~166--179, ACM, 2001.

\bibitem{rai2012zee}
A.~Rai, K.~K. Chintalapudi, V.~N. Padmanabhan, and R.~Sen, ``Zee: zero-effort
  crowdsourcing for indoor localization,'' in {\em Proceedings of the 18th
  annual international conference on Mobile computing and networking},
  pp.~293--304, ACM, 2012.

\bibitem{kontkanen2004topics}
P.~Kontkanen, P.~Myllymaki, T.~Roos, H.~Tirri, K.~Valtonen, and H.~Wettig,
  ``Topics in probabilistic location estimation in wireless networks,'' in {\em
  Personal, Indoor and Mobile Radio Communications, 2004. PIMRC 2004. 15th IEEE
  International Symposium on}, vol.~2, pp.~1052--1056, IEEE, 2004.

\bibitem{wu2004wlan}
C.-L. Wu, L.-C. Fu, and F.-L. Lian, ``Wlan location determination in e-home via
  support vector classification,'' in {\em Networking, sensing and control,
  international conference on}, vol.~2, pp.~1026--1031, IEEE, 2004.

\bibitem{prasithsangaree2002indoor}
P.~Prasithsangaree, P.~Krishnamurthy, and P.~Chrysanthis, ``On indoor position
  location with wireless lans,'' in {\em Personal, Indoor and Mobile Radio
  Communications, 2002. The 13th International Symposium on}, vol.~2,
  pp.~720--724, IEEE, 2002.

\bibitem{khan2014one}
S.~S. Khan and M.~G. Madden, ``One-class classification: taxonomy of study and
  review of techniques,'' {\em The Knowledge Engineering Review}, vol.~29,
  no.~03, pp.~345--374, 2014.

\bibitem{farkas2015placement}
K.~Farkas, ``Placement optimization of reference sensors for indoor tracking,''
  {\em Acta Polytechnica Hungarica}, vol.~12, no.~2, pp.~123--139, 2015.

\bibitem{baala2009impact}
O.~Baala, Y.~Zheng, and A.~Caminada, ``The impact of ap placement in wlan-based
  indoor positioning system,'' in {\em Networks, 2009. ICN'09. Eighth
  International Conference on}, pp.~12--17, IEEE, 2009.

\bibitem{bose2007practical}
A.~Bose and C.~H. Foh, ``A practical path loss model for indoor wifi
  positioning enhancement,'' in {\em Information, Communications \& Signal
  Processing, 2007 6th International Conference on}, pp.~1--5, IEEE, 2007.

\bibitem{scholkopf2001estimating}
B.~Sch{\"o}lkopf, J.~C. Platt, J.~Shawe-Taylor, A.~J. Smola, and R.~C.
  Williamson, ``Estimating the support of a high-dimensional distribution,''
  {\em Neural computation}, vol.~13, no.~7, pp.~1443--1471, 2001.

\bibitem{sun2010monotonicity}
Y.~Sun, {\'A}.~Baricz, and S.~Zhou, ``On the monotonicity, log-concavity, and
  tight bounds of the generalized marcum and nuttall $ q $-functions,'' {\em
  IEEE Transactions on Information Theory}, vol.~56, no.~3, pp.~1166--1186,
  2010.

\bibitem{howmany16}
O.~Georgiou, K.~Mimis, D.~Halls, W.~H. Thompson, and D.~Gibbins, ``How many
  wi-fi aps does it take to light a lightbulb?,'' {\em IEEE Access}, vol.~4,
  pp.~3732--3746, 2016.

\bibitem{nguyen2017wireless}
C.~L. Nguyen, O.~Georgiou, Y.~Yonezawa, and Y.~Doi, ``The wireless localisation
  matching problem,'' {\em IEEE Internet of Things Journal}, vol.~PP, no.~99,
  pp.~1--1, 2017.

\bibitem{powers2011evaluation}
D.~M. Powers, ``Evaluation: from precision, recall and f-measure to roc,
  informedness, markedness and correlation,'' 2011.

\end{thebibliography}
\end{document}